\documentstyle[12pt,epsfig]{article}
\def\lsim{\raise0.3ex\hbox{\,$<$\kern-0.75em\raise-1.1ex\hbox{$\sim$}\,}}
\def\gsim{\raise0.3ex\hbox{\,$>$\kern-0.75em\raise-1.1ex\hbox{$\sim$}\,}}

\begin{document}

\begin{titlepage}
\begin{flushright}
26 October, 2001\\
hep-ph/0110348
\end{flushright}
\begin{centering}
\vfill

{\bf NUCLEAR PARTON DISTRIBUTIONS IN THE DGLAP APPROACH}

\vspace{0.5cm}
 K.J. Eskola$^{\rm a,b,}$\footnote{kari.eskola@phys.jyu.fi},
 H. Honkanen$^{\rm a,}$\footnote{heli.honkanen@phys.jyu.fi}
 V.J. Kolhinen$^{\rm a,}$\footnote{vesa.kolhinen@phys.jyu.fi},\\
 P.V. Ruuskanen$^{\rm a,b,}$\footnote{vesa.ruuskanen@phys.jyu.fi} and
 C.A. Salgado$^{\rm c,}$\footnote{carlos.salgado@cern.ch}

\vspace{1cm}
{\em $^{\rm a}$ Department of Physics, University of Jyv\"askyl\"a,\\
P.O.Box 35, FIN-40351
Jyv\"askyl\"a, Finland\\}
\vspace{0.3cm}
{\em $^{\rm b}$ Helsinki Institute of Physics,\\
P.O.Box 64, FIN-00014 University of Helsinki, Finland\\}
\vspace{0.3cm}
{\em $^{\rm c}$ CERN, Theory Division, CH-1211 Geneva, Switzerland\\}

\vspace{1cm} \centerline{\bf Abstract} Determination of the nuclear
parton distributions within the framework of perturbative QCD, the
DGLAP equations in particular, is discussed.  Scale and flavour
dependent nuclear effects in the parton distributions are compared
with the scale and flavour independent parametrizations of HIJING and
of the Hard Probe Collaboration. A comparison with the
data from deep inelastic lepton-nucleus scattering and the Drell-Yan
process in proton-nucleus collisions is shown.
\end{centering}

\vfill
\end{titlepage}

\section{Introduction}

In a high-energy collision of hadrons or nuclei $A$ and $B$, inclusive
cross sections for the production of a particle $c$ involving a large
scale $Q\gg\Lambda_{\rm QCD}$, can in the leading twist approximation
be computed by using collinear factorization,
\begin{eqnarray}
& \displaystyle 
   d\sigma(Q^2,\sqrt s)_{AB\rightarrow c+X} =
   \sum_{i,j=q,\bar q,g} 
   \bigg[ Z_Af_i^{p/A}(x_1,Q^2)+(A-Z_A)f_i^{n/A}(x_1,Q^2)\bigg] 
   \otimes \nonumber  \\
 & \otimes\bigg[ Z_Bf_j^{p/B}(x_2,Q^2)+(B-Z_B)f_j^{n/B}(x_2,Q^2)\bigg]
   \otimes d\hat \sigma(Q^2,x_1,x_2)_{ij\rightarrow c+x}
%+ {\cal O}()
\label{hardAA}
\end{eqnarray}
where $d\hat \sigma(Q^2,x_1,x_2)_{ij\rightarrow c+x}$ is the
perturbatively calculable differential cross section for the
production of $c$ at the scale $Q$, $x_{1,2}\sim Q/\sqrt s$ are the
fractional momenta of the colliding partons $i$ and $j$, and
$f_i^{p(n)/A}$ is the distribution of parton flavour $i$ in a proton
(neutron) of the nucleus $A$, and correspondingly $f_j^{p(n)/B}$ is
that for the nucleus $B$.  The number of protons in $A(B)$ is denoted
by $Z_A (Z_B)$. In the leading twist approximation, on which we shall
focus in the following, multiple scattering of the bound nucleons
does occur but all collisions are independent of each other and only
one-parton densities are needed. At this level all the possible higher
twist terms, suppressed by $1/Q^2$ but enhanced by the nuclear
geometry (thickness of the nuclei), are neglected.  As correlations
between partons are not considered, the nuclear effects enter only via
the nuclear parton distribution functions (nPDF) $f_i^{p,n/A,B}$. The
nPDF differ from the parton distributions of the free proton but obey
the same DGLAP \cite{DGLAP} evolution equations. The DGLAP evolution
of the nPDF has been studied in e.g. \cite{QIU}-\cite{KUMANOnew}. For
studies of next-to-leading twist factorization involving two-parton
distributions, see \cite{QIU_ht,FRIES}.

The cross sections of hard processes measured in deeply inelastic $lA$
collisions and in $pA$ collisions offer the experimental constraints
necessary for pinning down the parton densities in nuclei. Once the
nPDF are known, the reference cross sections for hard probes of dense
matter in ultrarelativistic nucleus-nucleus collisions can be
computed and used in the  search of  the signals of the QGP. Another
motivation for the study of the nPDF is that the measurements of
certain hard processes in nuclear collisions are useful also for
extracting information of the parton distributions of the free proton,
provided that the contribution of the nuclear effects can first be
removed \cite{STIRLING}.

The purpose of this note is to discuss the constraints of the nuclear
parton densities within the DGLAP framework, such as the ones in
\cite{EKR,EKS98,KUMANOnew}. We shall compare the outcome of the studies
\cite{EKR,EKS98}, the EKS98 parametrization of the nuclear effects,
with scale and flavour independent parametrizations of HIJING \cite{HIJING} 
and the Hard Probe Collaboration \cite{HPC}, and especially, with the data.

\subsection{DIS and nPDF}

The cleanest way of getting information of the nPDF
is from deeply inelastic lepton-nucleus scattering (DIS) experiments. 
The differential cross section for deeply inelastic $lp$ scattering 
in the one-photon exchange approximation can be  expressed as 
\begin{equation}
\frac{d\sigma^{lp}}{dQ^2dx} = \frac{4\pi\alpha^2}{Q^4}
\frac{F_2^p(x,Q^2)}{x}\bigg\{
1-y-xyM/E_l+\frac{y^2}{2}\frac{1+4x^2M^2/Q^2}{1+R(x,Q^2)}
\bigg\},
\end{equation}
with the standard Lorentz-invariant variables $x=Q^2/(2p\cdot q)$ and
$y=p\cdot q/p\cdot k$, where $p$, $q$ and $k$ are the four-momentum of
the proton, the exchanged virtual photon, and the incoming lepton,
correspondingly.  Virtuality of the photon is $Q^2\equiv -q^2$, mass
of the nucleon is $M$, and $E_l$ is the energy of the incoming lepton
in the target rest frame.  In principle $0<x<A$ but for the discussion here
the small cumulative tails of the distributions at $x>1$ can be safely 
neglected.

From the measurements with different nuclear targets, one knows that
the ratio of the absorption cross sections of longitudinal and virtual
photons with the target nucleons, $R(x,Q^2)=
\sigma_L^{\gamma^*}/\sigma_T^{\gamma^*}$ does not significantly depend
on the target nucleus $A$ at scales $Q^2\gsim 1.5$ GeV$^2$
\cite{ARNEODO,HERMES}. The ratio of cross sections with different
targets thus reflects the ratio of the nuclear structure functions
$F_2^A/F_2^B$:
\begin{equation}
\frac{{d\sigma^{lA}}/{dQ^2dx}}
     {{d\sigma^{lB}}/{dQ^2dx}} 
\approx \frac{F_2^A(x,Q^2)}{F_2^B(x,Q^2)} 
= 
\frac{Z_AF_2^{p/A}(x,Q^2)+
(A-Z_A)F_2^{n/A}(x,Q^2)}{Z_BF_2^{p/B}(x,Q^2)+(B-Z_B)F_2^{n/B}(x,Q^2)},
\end{equation}
where $F_2^{p,n/A,B}$ are the structure functions of bound nucleons.
The ratio of the DIS cross sections from $lA$ and $lD$ 
is then related to the average structure functions per nucleon as
\begin{eqnarray}
R_{F_2}^A(x,Q^2)\equiv\frac{\frac{1}{A}{d\sigma^{lA}}/{dQ^2dx}}{\frac{1}{2}{d\sigma^{lD}}/{dQ^2dx}} 
\approx \frac{\frac{1}{A}F_2^A}{\frac{1}{2}F_2^D}
=\frac{\frac{1}{2}(F_2^{p/A} + F_2^{n/A})
+\frac{1}{2}(\frac{2Z}{A}-1)(F_2^{p/A} - F_2^{n/A})}
      {\frac{1}{2}(F_2^{p/D} + F_2^{n/D})},
\label{RF2}
\end{eqnarray}
where the numerator is 
written as a sum of isospin symmetric and non-symmetric terms.
The nuclear effects in deuterium are small, less than 1 percent, 
so to a first approximation these can be neglected.  Measurements of 
the ratio  $R_{F_2}^A(x,Q^2)$ have revealed clear and systematic 
nuclear effects in different regions of Bjorken-$x$ \cite{EMC}-\cite{E665-2}:
\begin{itemize}
\item ``shadowing''; a depletion  at $x \lsim 0.1$,
\item ``anti-shadowing''; an excess at $0.1 \lsim x \lsim 0.3$,
\item ``EMC effect''; a depletion at $0.3 \lsim x\lsim0.7$,
\item ``Fermi motion''; an excess towards $x\rightarrow1$ and beyond.
\end{itemize}
Systematics of $R_{F_2}^A$ in $A$ and in $x$ has been extensively
studied e.g. in the experiments EMC \cite{EMC,EMC-2,EMC-3} , SLAC
\cite{SLAC,SLACre}, BCDMS \cite{BCDMS},
NMC \cite{NMC95-1}-\cite{NMC96-2}, E665 \cite{E665-1,E665-2}. Since the $Q^2$
dependence of $R_{F_2}^A$ is quite weak at $x\gsim 0.1$ it has been more
difficult to probe. Data with high enough precision, however, exist:
the NMC has some years ago discovered a clear $Q^2$-dependence in the
ratio of the cross sections $d\sigma^{\mu{\rm Sn}}/d\sigma^{\mu{\rm C}}$ 
\cite{NMC96-2}, i.e. the scale dependence of the ratio 
$F_2^{\rm Sn}/F_2^{\rm C}$, at $x\sim 0.01$.

\section{The DGLAP analysis}

Generally, perturbative QCD (pQCD) cannot predict the absolute parton
distributions. However, once the starting distributions are given at a
scale $Q_0$, pQCD successfully predicts the evolution in
$Q^2$ in the form of the DGLAP evolution equations \cite{DGLAP}.  The
global DGLAP analyses of the parton distributions of the free proton,
such as MRS \cite{MRSnew} and CTEQ \cite{CTEQ5L}, involve a
determination of those input distributions which, when evolved to
different (higher) values of $Q$, give the best overall agreement with
the data from different hard processes. Conservation of momentum and
baryon number are maintained by the DGLAP evolution, and they are used
as further constraints.

In the leading twist framework the situation is exactly the same
for the nPDF: the data from hard processes in
nuclear collisions at various values of $x$ and $Q$, together with
the momentum and baryon number conservation, can be used to constrain
the nonperturbative input distributions of partons in bound protons at
some initial scale $Q_0\gg \Lambda_{\rm QCD}$. The link between the scales 
$Q_0$ and $Q$ is given by the DGLAP equations in the whole range of 
$x$ studied. 

In general, the perturbative QCD scale evolution of the nPDF has been
extensively studied in the literature, see e.g.
\cite{QIU}-\cite{KUMANOnew} and \cite{SARCEVIC,AYALA}. The origin of
the nuclear effects is an interesting question but is beyond the scope
of the DGLAP analysis. It has been suggested that for the DGLAP
evolution one may compute the initial conditions at a scale $Q_0$
from a model, as is done e.g. in \cite{STRIKMAN, KUMANO}, and then apply
the DGLAP equations to describe the evolution in $Q$.  Nevertheless,
even in this case the key feature is the detailed comparison with the
existing data. DGLAP analyses of the nPDF which attempt to rely
only on the data in the determination of the initial conditions, are
presented in Refs. \cite{EKR,EKS98,KUMANOnew}.

\subsection{Quarks and antiquarks}

In the QCD-improved parton model (in leading order, or in the DIS-scheme 
in any higher order) the structure function $F_2$ of the proton(neutron) 
can be written in terms of its parton distributions as
\begin{equation}
F_2^{p(n)/A}(x,Q^2)= \sum_{q=u,d,s,...}
e_q^2\bigg[xf_q^{p(n)/A}(x,Q^2)+xf_{\bar q}^{p(n)/A}(x,Q^2)\bigg].
\end{equation}
As in the case of the free nucleons, the parton distributions of bound
neutrons in isoscalar nuclei are obtained through isospin symmetry,
$f_{u(\bar u)}^{n/A}=f_{d(\bar d)}^{p/A}$ and $f_{d(\bar
d)}^{n/A}=f_{u(\bar u)}^{p/A}$.  This is expected to be a good
approximation for non-isoscalar nuclei as well.

It is convenient to define the nPDF for each parton flavor $i$ through
the modifications of the corresponding distributions in the free
proton,
\begin{equation}
R_i^A(x,Q^2)\equiv {f_i^{p/A}(x,Q^2)\over f_i^p(x,Q^2)},
\label{eqratios}
\end{equation}
where we shall assume that the parton distributions of the free
proton are fully known. For example, below the mass threshold of the charm
quark, we can write
\begin{equation}
R_{F_2}^A(x,Q^2) = \frac{5(u_A+\bar u_A + d_A +\bar d_A) + 4s_A
+(\frac{2Z}{A}-1)3(u_A+\bar u_A - d_A-\bar d_A)}
     {5(u+\bar u + d +\bar d) + 4s},
\label{RF20}
\end{equation}
where $u_A \equiv f_u^{p/A} = R_u^A(x,Q^2)f_u^p(x,Q^2)$, and similarly
for the other quarks.  The ratio $R_{F_2}^A$ measured at these scales thus
constrains the individual ratios $R_q^A$ and $R_{\bar q}^A$ in certain
combination.

Obviously, more constraints are needed in order to pin down the ratios
$R_i^A$.  These can be obtained from the measurements of the Drell-Yan
dileptons in $pA$ collisions by E772 \cite{E772} and E866 \cite{E866}
which offer high enough precision for getting statistically
significant constraints. In the lowest order, the ratio of the
differential cross sections for the Drell-Yan process in $pA$ and $pD$
collisions is given by
\begin{eqnarray}
&\!\!\!\!\!R&\!\!\!\!\!_{DY}^A(x_2,Q^2) \equiv  \frac
{\frac{1}{A} {d\sigma^{pA}_{DY}}/{dx_2dQ^2}}
{\frac{1}{2} {d\sigma^{pD}_{DY}}/{dx_2dQ^2}}
\nonumber\\
&&\!\!\!\!\!\!\!
 = \{4[u_1(\bar u_2^A+\bar d_2^A)+ \bar u_1(u_2^A+d_2^A)] +
  [d_1(\bar d_2^A+\bar u_2^A) + \bar d_1(d_2^A+u_2^A)] +
4s_1s_2^A +...\}/N_{DY}  \nonumber\\
&&\!\!\!\!\!\!\!+ (\frac{2Z}{A}-1)
\{4[u_1(\bar u_2^A-\bar d_2^A) + \bar u_1(u_2^A-d_2^A)]+
[d_1(\bar d_2^A-\bar u_2^A) +  \bar d_1(d_2^A-u_2^A)]\}/N_{DY}
\label{RDY}
\end{eqnarray}
where the invariant mass of the lepton pair is $Q^2$, and 
the subscript 2 (1) refers to the fractional momentum $x_2$ ($x_1$)
of the parton from the target (proton). The denominator is 
\begin{equation}
N_{DY} = 4[u_1(\bar u_2+\bar d_2)+\bar u_1(u_2+d_2)] +
  [d_1(\bar d_2+\bar u_2) +\bar d_1(d_2+u_2)] + 4s_1s_2 + ...
\end{equation}
with the dots denoting the heavier flavours.  Again, the ratios
$R_q^A$ and $R_{\bar q}^A$ are probed in certain combination.  The
typical $x$ and $Q^2$ range probed by the measurements of the ratios
$R_{F_2}^A$ and $R_{DY}^A$ can be seen in Fig. 1.

\begin{figure}[hbt]
\vspace{-2.5cm}
%\centerline{\epsfxsize=10cm\epsfbox{/n/qgp/pub/HH/KUVAT/x_vs_Q2.eps}}
\centerline{\epsfxsize=10cm\epsfbox{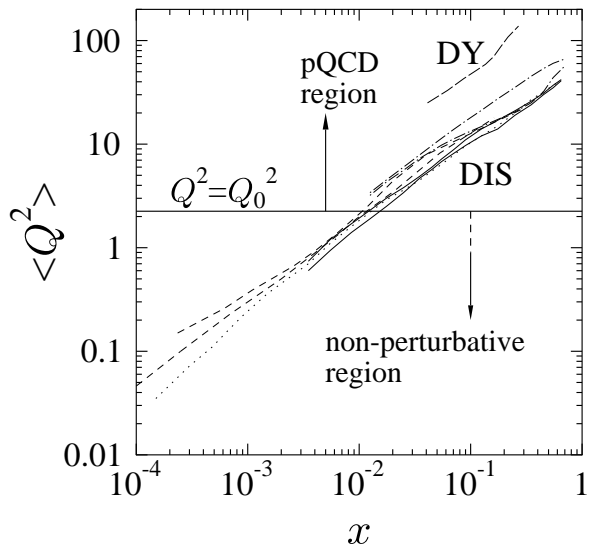}}
\vspace{-2.5cm}
\caption[a] {\small The correlation between the Bjorken 
$x$ and the photon virtuality $Q^2$ in deep inelastic $\mu A$
scattering measured by the NMC \cite{NMC95-1,NMC95-2,NMC96-1} and E665
\cite{E665-1,E665-2} (denoted by DIS). The same (with $x=x_2$) for the
measurements of the Drell-Yan process in $pA$ by E772 \cite{E772}
(denoted by DY). The starting scale of the DGLAP analysis \cite{EKR}
is $Q_0^2$.  }
\label{x_vs_Q}
\end{figure}

In addition, conservation of baryon number, 
\begin{eqnarray} 
3 &=& \int_0^A dx \sum_{q=u,d}\bigg[f_q^{p/A}(x,Q^2)-f_{\bar q}^{p/A}(x,Q^2)\bigg]\cr 
&\approx&  \int_0^1 dx
\sum_{q=u,d}\bigg[R_q^A(x,Q^2)f_q^p(x,Q^2)-R_{\bar q}^A(x,Q^2) f_{\bar q}^p(x,Q^2)\bigg],
\label{baryon}
\end{eqnarray}
can be used to pin down the valence quark distributions \cite{STRIKMAN,KJE}. 
On the r.h.s. of Eq. (\ref{baryon}), the small cumulative tails at $x>1$  
have been neglected.

Even in an ideal case, where the experimentally measured DIS and DY
ratios of Eqs. (\ref{RF2}) and (\ref{RDY}) would lie along a constant
scale $Q_0$ in a wide range of $x$, the number of experimental
constraints above would not be enough to fully fix the ratios
$R_q^A(x,Q_0^2)$ and $R_{\bar q}(x,Q^2)$. In reality, the situations
is even more difficult: the data on both DIS and DY are
given only in distinct regions, within which the values of $x$ are
strongly correlated with $Q^2$, as illustrated in Fig. \ref{x_vs_Q}.
Therefore, a recursive procedure similar to that in the global
analyses of the parton distributions of the free proton has to be
adopted in order to determine the initial ratios $R_i^A(x,Q^2_0)$. 
Notice that in comparison with the free proton case, an additional 
variable, the mass number $A$, appears.

To simplify the determination of the input nuclear effects for valence
and sea quarks (without invoking any specific model) one may in a
leading approximation assume them to be separately flavor-independent:
$R_{u_V}^A(x,Q_0^2)\approx R_{d_V}^A(x,Q_0^2) \approx R_V^A(x,Q_0^2)$,
and $R_{\bar u}^A(x,Q_0^2)\approx R_{\bar d}^A(x,Q_0^2) \approx
R_{s}^A(x,Q_0^2)\approx R_S(x,Q_0^2)$ \cite{KJE, EKR}. Note that this
approximation is needed {\em only} at $Q_0^2$ but the observation in
\cite{EKR} is that it remains good also in the evolution to higher
$Q^2$. In this approximation the problem reduces to constraining
the three independent input ratios, $R_V^A$, $R_S^A$ and $R_G^A$ at
the initial scale $Q_0^2$. The details of an analysis using this approach
can be found in \cite{EKR}, below we only summarize the available 
constraints in each region of $x$.

\begin{itemize}
\item At { $x\gsim 0.3$} the valence quark distributions dominate the
structure function $F_2^A$, and $ R_{F_2}^A\approx R_V^A$.  The DIS
data for $R_{F_2}^A$ therefore only constrains the magnitude of the
EMC effect and the Fermi-motion in $R_V^A$ but not in $R_S^A$ or in
$R_G^A$. For the sea quarks, it is assumed in \cite{EKR} that
$R_S(x\gsim0.3,Q_0^2)\approx R_V(x\gsim0.3,Q_0^2)$ (cf. the discussion
for gluons below). To what extent the Drell-Yan production measurable in
nuclear collisions at the SPS could probe the EMC effect of the
nuclear sea quarks, was recently studied in Ref. \cite{EKST}.

\item At {$0.04\lsim x \lsim 0.3$} the DIS and DY data both constrain
the ratios $R_S^A$ and $R_V^A$ but from different regions of $Q^2$, as
shown in Fig. \ref{x_vs_Q}. In addition, $R_V^A$ is restricted by 
conservation of baryon number. An outcome of the DGLAP analyses
\cite{STRIKMAN,KJE,EKR} is that at the input scale $Q_0^2\sim 2$
GeV$^2$ the sea quark content of a nucleus remains smaller than that
of the free proton, i.e. that no antishadowing appears in
$R_S^A(x,Q_0^2)$.

\item At { $x\lsim0.04$} the DIS data for the ratio $R_{F_2}^A$ extend
down to $x\sim 5\cdot10^{-3}$ in the region $Q\gsim1$ GeV relevant for
the DGLAP analysis.  In the analyses \cite{KJE,EKR,KUMANOnew} the nuclear
valence quarks have less shadowing than the sea quarks, which is
mainly due to the conservation of baryon number. Also more strongly
shadowed valence quarks have been suggested \cite{STRIKMAN}.

\item The DIS data for $R_{F_2}^A$ at {\bf $x\lsim 5\cdot10^{-3}$}
only exist in the region $Q\lsim1$ GeV which can be considered
nonperturbative and not treatable with the DGLAP equations (see
Fig. \ref{x_vs_Q}).  A saturation behaviour, flattening of $R_{F_2}$
in $x \rightarrow 0$, is observed along the experimentally probed
values of $Q^2$ \cite{NMC95-2,E665-1}.  Such a behaviour, a weak
dependence of $R_{F_2}^A$ on $x$, can also be expected at $Q_0^2$,
provided that the sign of the slope of the $Q^2$ dependence of
$R_{F_2}^A$ in the nonperturbative region remains the same (positive)
as what is measured at $x\sim 0.01$ in the perturbative region
\cite{NMC96-2}.  The DIS data in the non-perturbative region should
thus give a lower bound for $R_{F_2}^A(x,Q_0^2)$ at small values of
$x$.  The sea quarks dominate over the valence quarks in this region,
so practically only the ratio $R_S^A$ is constrained by the DIS data.
\end{itemize}

\subsection{Gluons}

For the gluons the situation is less straightforward since direct
measurements of the gluon distributions in nuclei are difficult.  In
principle the measurements of $D$-mesons in $pA$ collisions at various
cms-energies can be used for pinning down the nuclear gluon
distributions \cite{ZIWEI,KOLHINEN}. These measurements already exist
but so far the error bars in the results of experiment E789
\cite{DinpA} are too large for getting a stringent constraint for
$R_G^A$.  The future measurements of $D$ production in $pA$ collisions
by the NA60 experiment at SPS, by PHENIX at RHIC and hopefully also by
ALICE or CMS at the LHC, will provide very important input for fixing
the gluon distributions in nuclei \cite{KOLHINEN}.  Also direct
photons in nuclear collisions can be used for this purpose,
possibly also diffractive scattering in DIS \cite{STRIKMAN_diff}.
Production of $J/\Psi$ in $pA$ always involves strong final state
effects ($J/\Psi$ suppression in normal nuclear matter), which makes
the extraction of the initial state effects from the data very
difficult \cite{SALGADO}.

In the DGLAP analysis, the best (but still indirect) constraint  on the
ratio $R_G^A(x,Q^2)$ is provided by the measurements by NMC
\cite{NMC96-2} of the $Q^2$ dependence of the ratio $F_2^{\rm
Sn}/F_2^{\rm C}$. The scale evolution of $F_2$ is coupled to the gluon
distributions at small values of $x$, where gluons dominate, 
approximately as \cite{PRYTZ}
\begin{equation}
{\partial F_2(x,Q^2)}/{\partial \log Q^2} \approx
\frac {10}{27} \frac{\alpha_s}{\pi} xg(2x,Q^2).
\label{PRYTZ}
\end{equation}

Based on Eq. (\ref{PRYTZ}), the $Q^2$ slope of $R_{F_2}^A$ at small
values of $x$ becomes
\begin{equation}
\frac{\partial R_{F_2}^A(x,Q^2)}{\partial \log Q^2}
\approx
\frac{10\alpha_s}{27\pi}\frac{xg(2x,Q^2)}{F_2^D(x,Q^2)}
\biggl\{R_G^A(2x,Q^2)-R_{F_2}^A(x,Q^2)\biggr\},
\label{RF2slope}
\end{equation}
which suggests that the more deeply gluons are shadowed, the slower is
the evolution of $R_{F_2}^A$.  So far only the NMC data \cite{NMC96-2}
of the $Q^2$ dependence of the ratio $F_2^{\rm Sn}/F_2^{\rm C}$ has
sufficient precision for getting stringent constraints, as first
pointed out in Ref. \cite{PIRNER}.

Yet another indirect constraint, the momentum sum rule, 
\begin{equation}
1
=       \int_0^A dx \sum_{i=g,u,\bar u,...} xf_i^{p/A}(x,Q^2)
\approx \int_0^1 dx \sum_{i=g,u,\bar u,...} R_i^A(x,Q^2)xf_i^{p}(x,Q^2)
\label{momentum}
\end{equation}
should be taken into account in the DGLAP approach. A few percent
flow of momentum from quarks and antiquarks to gluons is expected
relative to the free proton case \cite{STRIKMAN, KJE}. Since 
$\int_0^1 dx R_G^A(x,Q_0^2)xg_p(x,Q_0^2) > \int _0^1 dx\, xg_p(x,Q_0^2)$, 
antishadowing  $R_G^A(x,Q^2_0)> 1$ is bound to exist in some region of
$x$. Below we summarize the constraints on $R_G^A(x,Q_0^2)$ in
different regions of $x$.

\begin{itemize}
\item At $0.02 \lsim x \lsim0.2$ the NMC data on the $Q^2$ evolution
of the ratio $F_2^{\rm Sn}/F_2^{\rm C}$ \cite{NMC96-2} set the main
constraint through the DGLAP evolution (the full equations;
Eq. (\ref{RF2slope}) holds at the small values of $x$ only). These NMC
data extend down to $x=0.0125$, so only gluons at $x\gsim 0.02$ can be
constrained. Especially, it is observed that the $Q^2$ slope of the
ratio $F_2^{\rm Sn}/F_2^{\rm C}$ is clearly positive at small values
of $x$, indicating that obviously also the $Q^2$ slope of
$R_{F_2}^A(x,Q^2)$ is positive. Then, according to Eq.
(\ref{RF2slope}), $R_G^A(2x,Q^2)>R_{F_2}^A(x,Q^2)$. The data thus seems
to rule out the case where the shadowing of nuclear gluons is much
stronger than the shadowing observed in $R_{F_2}^A$. In a solution
consistent with the data on $F_2^{\rm Sn}/F_2^{\rm C}$ {\em and} with
the momentum sum rule the gluons have less shadowing than the sea
quarks and more antishadowing than the valence quarks or in
$R_{F_2}^A$ \cite{EKR,PIRNER}.

\item At $x\lsim0.02$ stringent experimental constraints do not exist
for the gluons at the moment. Assuming, however, that the $Q^2$ slope
of $R_{F_2}^A$ remains positive, the measured saturation of shadowing
in $R^A_{F_2}$ at $Q^2\ll 1$ GeV$^2$ \cite{NMC95-2,E665-1} gives the
lower bound for the shadowing of $R^A_{F_2}(x,Q_0^2)$. The weak $x$
dependence of $R_{F_2}^A$ (together with $\partial R_{F_2}^A/\partial
Q^2 > 0$) indicates also a weak $x$ dependence of $R_G^A(x,Q_0^2)$, and
it is then concievable to expect that $R_G^A(x,Q_0^2)\approx
R_{F_2}^A(x,Q_0^2)$ for $x\ll1$ \cite{EKR}. The evolution is stable in
the sense that the approximate equality remains true within about 5 \%
even after the DGLAP evolution from $Q_0\sim 1$ GeV to $Q\sim100$ GeV
\cite{EKS98}. In order to pin down the gluons in this region,
high-precision DIS measurements at small values of $x$ (but $Q^2\gsim
1$ GeV$^2$) would be needed. More constraints in this region of $x$
are expected from the measurements of dileptons originating from the
decays of the $D$ mesons in PHENIX at RHIC \cite{ZIWEI, KOLHINEN} and
in ALICE or CMS at the LHC \cite{KOLHINEN}.

\item At $x\gsim0.2$ there are currently no clear experimental constraints
available for the gluons. Conservation of momentum does not reveal 
whether an EMC effect exists for the
gluons or not: about 30 \% of the gluon momentum comes from $x\gsim0.2$,
so, say, a 10 \% net change in the momentum content of the EMC region
can be compensated by roughly a 6\% net effect in the region where the
antishadowing bump is anticipated and which contains about half of the
gluon momentum.  Thus, the amount of antishadowing is not affected to
the extent that it would violate the constraints obtained from $x\lsim
0.2$ (see the estimates of the uncertainties in \cite{PIRNER}).
In the DGLAP evolution equations, the valence quarks act as source of
gluons in the gluon evolution, and the gluons in turn feed the sea
quark evolution. In course of the scale evolution, the experimentally
verified EMC effect of valence quarks will generate a similar EMC
effect for the gluons which in turn transmit the effect into the sea
quark distributions, as seen in \cite{KJE}, where no input EMC effect
was assumed for the gluons. A consistent assumption in the DGLAP
framework therefore is to include an EMC effect already for the
initial ratios $R_G^A$ and $R_S^A$. In this way the nuclear
modifications $R_i^A$ remain stable against the evolution, i.e. they
do not rapidly evolve away from their input values.  In the future,
however, experimental constraints for the nuclear gluons in this
region of $x$ could be obtained from the measurements of dileptons 
(originating from the decays of the $D$ mesons) by the NA60 experiment 
at the SPS \cite{KOLHINEN}.

\end{itemize}

\subsection{The EKS98 parametrization}

We have discussed above how to get constraints for the nPDF.  In the
DGLAP framework \cite{EKR} the problem boils down to determining
the input distributions, i.e. the input ratios $R_G^A(x,Q_0^2)$,
$R_V^A(x,Q_0^2)$ and $R_S^A(x,Q_0^2)$.  In practise, the extraction of
the input ratios involves a recursive procedure: first the DGLAP
evolution is performed with some input distributions, then a
comparison with the data is made at various values of $x$ and $Q^2$,
after which the input distributions are changed in order to achieve a
better agreement with the data.  This iterative procedure is repeated
until the best set of $R_G^A(x,Q_0^2)$ $R_V^A(x,Q_0^2)$ and
$R_S^A(x,Q_0^2)$ is found.  The details can be found in \cite{EKR}.

It is clear that the constraints always restrict the absolute 
nPDF $f_i^{p/A}$ (see e.g. Eqs. (\ref{RF2}) and (\ref{RDY})).
Constraints for the ratios $R_i^A$ depend on the chosen set of
parton distributions of the free proton, in terms of  which the ratios
in Eq. (\ref{eqratios}) are defined.  Ideally of course there exists only
one best set $\{ f_i^p\}$, but in practise, as some uncertainties
still appear also in the free proton level, several sets are in use,
(MRS, CTEQ, GRV, etc.). To test how strongly the obtained nuclear
effects might depend on the choice of the parton distributions of the
free proton, we have repeated the analysis of \cite{EKR} in
\cite{EKS98} by using the CTEQ4L \cite{CTEQ94} distributions instead
of the GRVLO distributions \cite{GRVLO}. In spite of the quite large
differences in the gluon content (at small $x$ there are fewer gluons
in CTEQ4L) and in the sea quarks (more flavour asymmetry in CTEQ4L)
between these sets, the differences between $R_{i,{\rm GRV}}^A$ and
$R_{i,{\rm CTEQ}}^A$ remained within a few percents only. Accepting
this range of uncertainty, it is meaningful to prepare a ``universal''
parametrization for the ratios $R_i^A(x,Q^2)$.  This task was
performed in \cite{EKS98}, and the parametrization, called ``EKS98'',
is available for public use in the www \cite{EKS98loc} and now also
in the latest version of the CERN PDFLIB \cite{PDFLIB}.

\section{Comparison of different parametrizations}

Next we shall compare the EKS98-parametrization with two other
parametrizations of the nuclear effects used in the literature, the
default one in HIJING \cite{HIJING} and the one prepared in one of the
Hard Probe Workshops \cite{HPC}. Especially, we shall focus on the
comparison of all these against the data.

\subsection{The HIJING parametrization}

The HIJING parametrization of the nuclear modifications of the parton 
distributions given in Ref. \cite{HIJING} is the following:

\begin{equation}
R_{F_2}^A(x) = 1+1.19(\ln A)^{1/6}[x^3-1.5(x_0+x_L)x^2+3x_0x_Lx] 
 -[\alpha_A-\frac{1.08(A^{1/3}-1)}{\ln(A+1)}\sqrt x] {\rm e}^{-x^2/x_0^2},
\end{equation}
where $\alpha_A=0.1(A^{1/3}-1)$, $x_0=0.1$ and $x_L=0.7$.  The
parametrization is based on fits to the old EMC data on $R_{F_2}^A$
\cite{EMC-2}. In particular, in the applications of this
parametrization, it is assumed that the modifications are identical
for all parton flavours, and that the $Q^2$ dependence of the ratios
is a negligible effect, $R_i^A(x,Q^2)=R_{F_2}^A(x)$.

\subsection{The HPC parametrization}

An outcome of the Hard Probe Collaboration (HPC) meeting in ECT$^*$ at
Trento (1995), was a parametrization \cite{HPC} of
$R_{F_2}^A(x)$. Similarly to the HIJING parametrization above, this
parametrization assumes that the $Q^2$ dependence is a negligible
effect. Possible differences between the modifications of different
parton species are not considered, either. The HPC fit was motivated 
by a corresponding parametrization in \cite{VARY} and it
was obtained by fitting the re-analysed SLAC data
\cite{SLACre} and re-analyzed NMC data \cite{NMC95-1}.  
The functional form of the HPC fit is

\begin{equation}
R_{F_2}^A(x) = 
\left\{
\begin{array}{ll}
R_{\rm sh}\frac{\displaystyle 1+c_Dc_A(1/x-1/x_{\rm sh})}{\displaystyle 1+c_A
A^{p_A}(1/x-1/x_{\rm sh})}, & \mbox{$x\le x_{\rm sh}$} \\ \\ 
a_{\rm emc}-b_{\rm emc}x, & \mbox{$ x_{\rm sh}\le x\le x_{\rm f}$} \\ \\ 
R_f\bigg(\frac{\displaystyle 1-x_f}{\displaystyle 1-x}\bigg)^{p_{\rm f}}, 
				& \mbox{$x_{\rm f} \le x\le 1$}
\end{array}
\right.
\end{equation}
where the different regions are matched together by setting
$R_{\rm sh}=a_{\rm emc}-b_{\rm emc}x_{\rm sh}$ and 
$R_f = a_{\rm emc}-b_{\rm emc}x_f$.  
In the EMC region 
$a_{\rm emc} = 1 + b_{\rm emc}x_{\rm emc}$. 
The $A$ dependence of $b_{\rm emc}$ is  
%\begin{equation}
$b_{\rm emc}= p_{\rm emc}[ 
1-A^{-1/3}-1.145 A^{-2/3}+0.93 A^{-1} + 0.88 A^{-4/3}-0.59 A^{-5/3}]
$
%\end{equation}
from  Ref. \cite{SMIRNOV}.
It was assumed (based on the data and the assumed $Q^2$ independence) 
that $R_{F_2}^A(x_{\rm emc})=1$ and that the location of the EMC minimum
is at $x_f$ independently of $A$.  
The eight fit parameters obtain the following values: 
%in the shadowing region
$p_A= 0.10011$, $c_A=0.0127343$, $c_D = 1.05570$ $x_{\rm sh}  = 0.154037$,
%In the region of the EMC effect 
$x_{\rm emc} = 0.275097$, $p_{\rm emc} = 0.525080$,
%In the region of Fermi motion 
$x_{\rm f} = 0.742059$ and $p_{\rm f} = 0.320992$.

\begin{figure}[thb]
\vspace{-1cm}
%\centerline{\epsfxsize=11cm\epsfbox{/n/qgp/pub/HH/KUVAT/rf208.eps}\hspace{9cm}}
\centerline{\epsfxsize=11cm\epsfbox{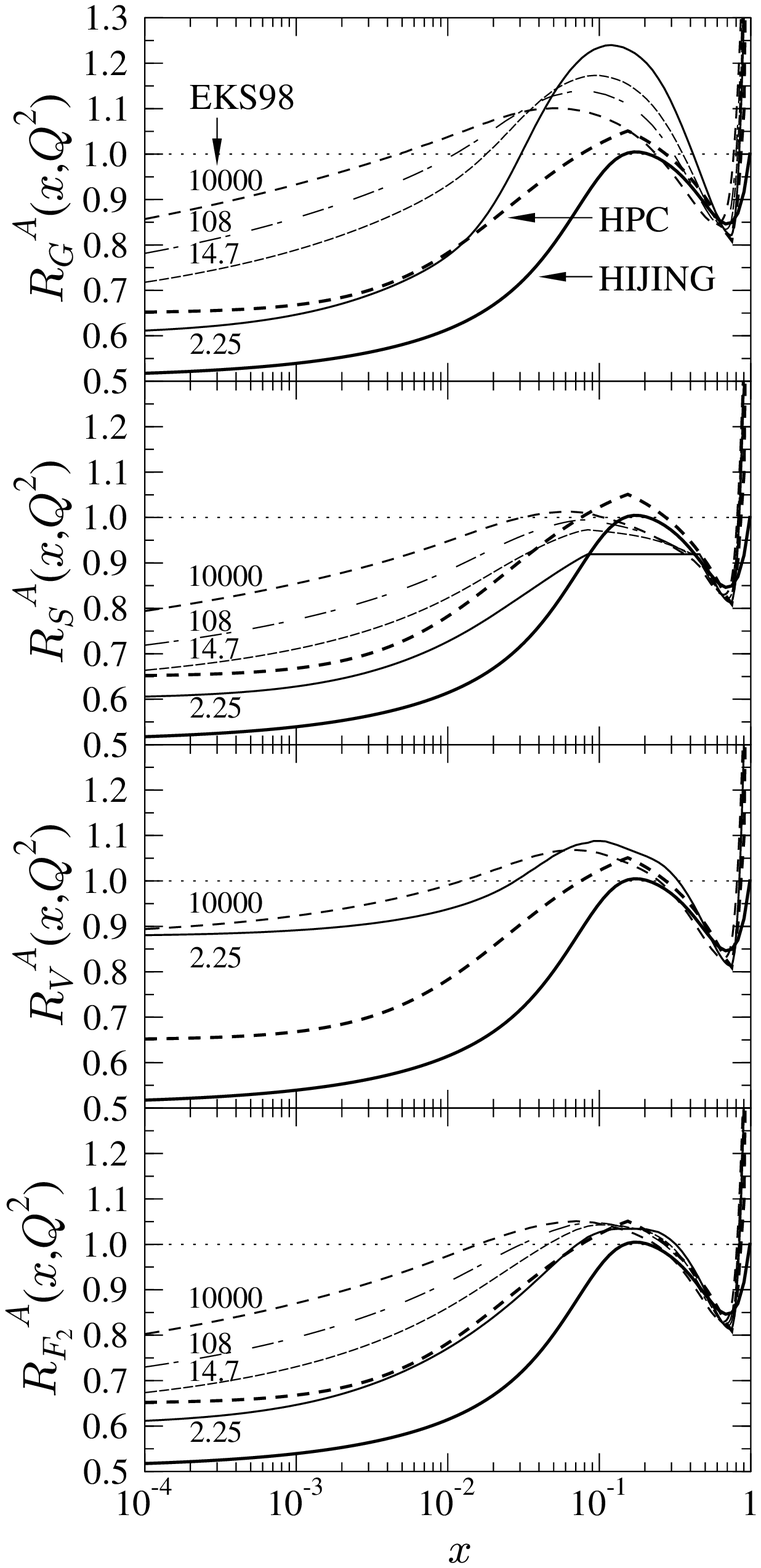}\hspace{9cm}}
\vspace{-10cm}\hspace{8cm}
\begin{minipage}{7cm}
%\caption{
\small Figure 2.  The nuclear modifications of parton densities for a
heavy nucleus $A=208$ (isoscalar). The ratios $R_G^A(x,Q^2)$,
$R_S^A(x,Q^2)$, $R_V^A(x,Q^2)$, and $R_{F_2}^A(x,Q^2)$ from
Ref. \cite{EKR} are denoted by EKS98, (thin lines, plotted at the
fixed values of $Q^2/$GeV$^2$ indicated on the left).  The $Q^2$ independent
parametrizations of the ratio $ R_{F_2}^A(x)$ of HIJING (thick solid
line) and that of HPC (thick dashed line) remain unchanged from panel 
to panel.  The notation  is the same  in all panels.
\end{minipage}
%}
\vspace{4cm}
\end{figure}

\begin{figure}[tb]
\vspace{-1.5cm}
\centerline{
\epsfxsize=15cm\epsfbox{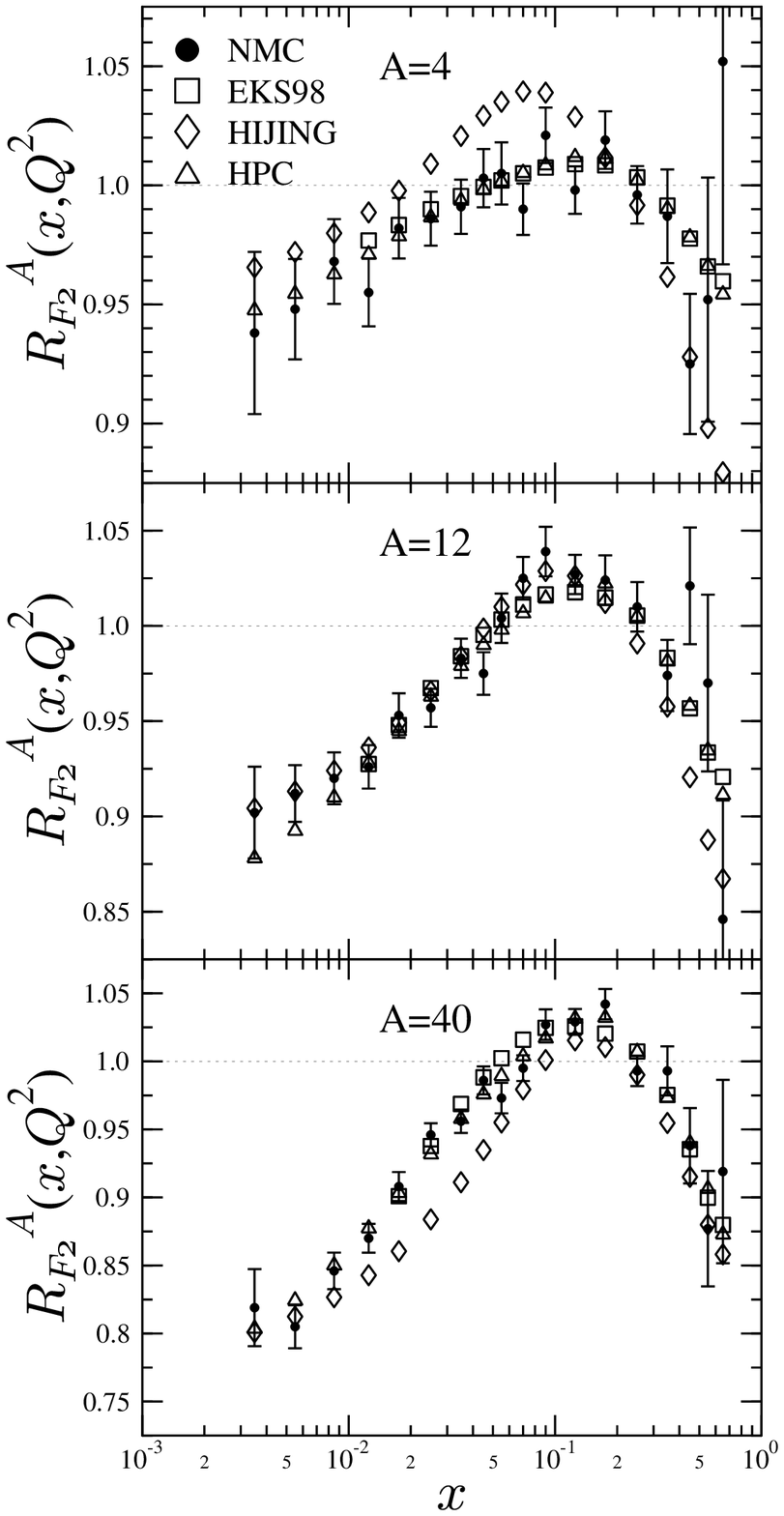}\hspace{5cm}}
\vspace{-15cm}\hspace{8cm}
\begin{minipage}{7cm}
\small Figure 3.  Comparison of the EKS98 (open squares), HIJING (open
diamonds) and HPC (open triangles) parametrizations with the NMC data
(filled circles with error bars) for $F_2^A/F_2^D$ for helium, carbon
and calcium \cite{NMC95-1} . For EKS98 the comparison is only made for
the experimental points above $Q^2=2.25$ GeV$^2$. Notice that the
vertical scales are different in each panel.
\end{minipage}
\vspace{7cm}
\end{figure}

\subsection{The Comparison}

The results of the DGLAP analysis \cite{EKR} for the ratios
$R_G^A(x,Q^2)$, $R_S^A(x,Q^2)$, $R_V^A(x,Q^2)$, and $R_{F_2}^A$ at
scales $Q^2=$ 2.25 GeV$^2$ (thin solid lines), 14.7 GeV$^2$ (dashed),
108 GeV$^2$ (dotted-dashed) and 10000 GeV$^2$ (dashed) are compared
with the HIJING parametrization (thick solid) and with the HPC
parametrization (thick dashed) in Fig. 2. This figure is to illustrate
two points: first, the $Q^2$ dependence becomes a non-negligible
effect at small values of $x$.  Secondly, the available experimental
and sum-rule constraints lead to mutually quite different
modifications for the valence quarks, sea quarks and gluons. For
$A=208$, the HIJING parametrization underestimates the baryon number
sum rule by 18~\% (24~\%) and the momentum sum rule by 13~\% (23~\%)
at $Q^2=Q_0^2=2.25$ GeV$^2$ ($Q^2=10^4$ GeV$^2$).  For the HPC
parametrization the corresponding figures are 5~\% (12~\%) for the
baryon number and 8~\% (12~\%) for the momentum sum rule.  The effects
of choosing different (lowest order) free proton parton distributions
remain within one unit of percent.  The deficit can easily be
understood from the figure by comparing with the EKS98 in which these
conservation laws are met within an accuracy of $\sim$ 1~\% at all
$Q^2$ for any (lowest order) PDF set of the free proton.

Fig. 3 shows the comparison of the EKS98 (open squares), HIJING (open
diamonds) and HPC (open triangles) parametrizations with the NMC data
\cite{NMC95-1} for $R_{F_2}^A$ (filled circles). The EKS98 results are
computed by using CTEQ5L set of parton distributions
\cite{CTEQ5L,PDFLIB} and at scales $Q^2$ corresponding to the $\langle
Q^2\rangle$ measured for each $x$ (see Fig. 1). The differences remain
small between the EKS98, HPC and the data.  HIJING agrees with the
data for carbon but overestimates the $A$-dependence of shadowing for
other nuclei. The error bars in the data represent the statistical and
systematic errors added in quadrature.

\begin{figure}[hbt]
\vspace{-1.5cm}
\centerline{
\epsfxsize=13cm\epsfbox{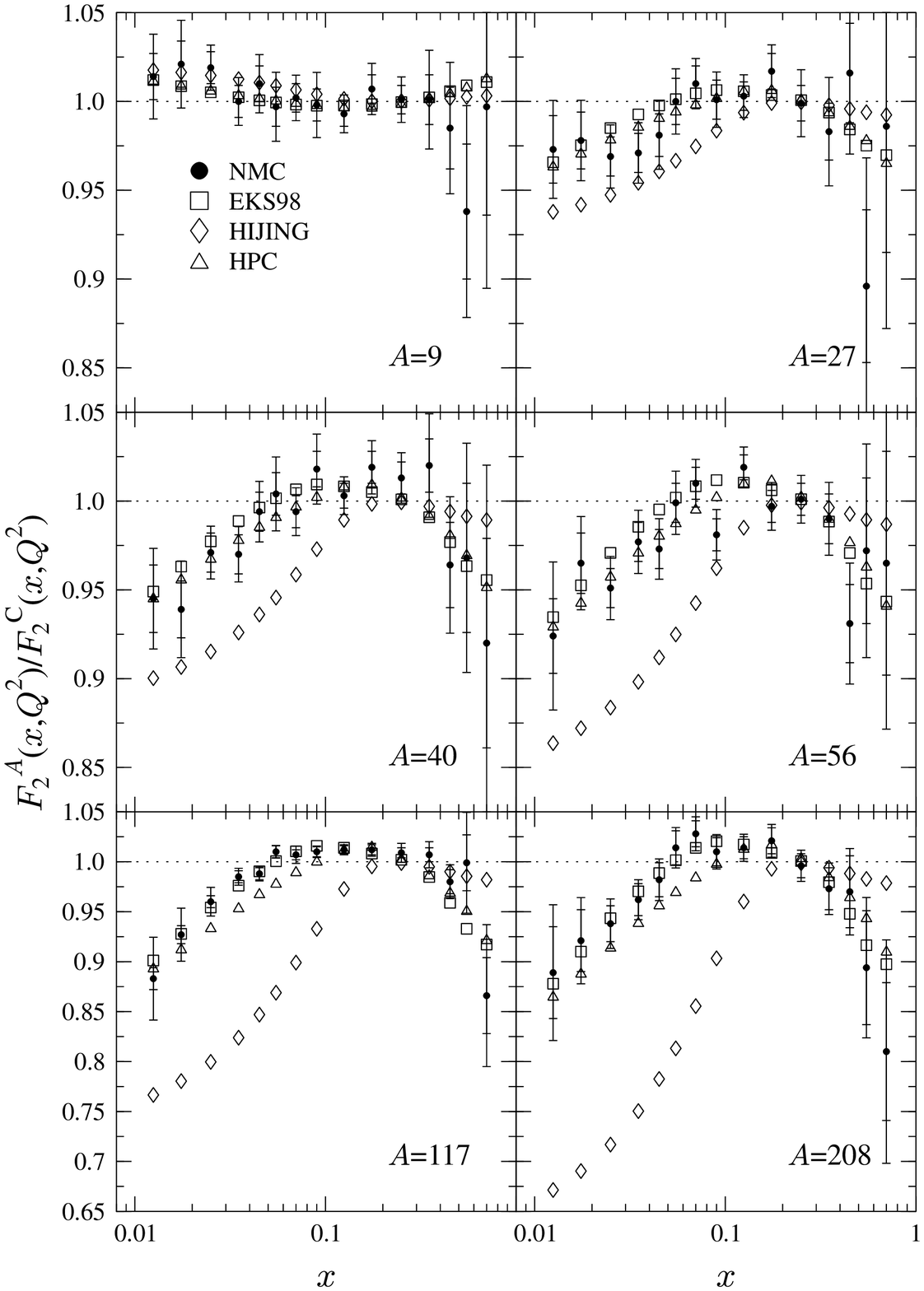}}
\vspace{0cm}
{
\small   Figure 4. Comparison of the EKS98 (open squares), HIJING (open
diamonds) and HPC (open triangles) parametrizations with the NMC data
(filled circles with error bars) for $F_2^A/F_2^{\rm C}$ \cite{NMC96-1}.
}
\vspace{-0.5cm}
\end{figure}

Next, in Fig. 4 we show the comparison of EKS98, HIJING and HPC with
the $A$ systematics of $R_{F_2}^A$ measured by the NMC \cite{NMC96-1}.
Again, the data is shown by the filled circles with statistical errors
(the inner error bars) and with the statistical and systematic errors
added in quadrature (the outer error bars). The notation is the same
as in the previous figure, and the EKS98 results are again computed at
the scales $Q^2=\langle Q^2\rangle$ for each value of $x$ given by the
experiment.  This figure shows that the $A$ dependence of shadowing is
clearly too strong in HIJING. It should also be noted that the
data of Fig. 4 was not yet available for the HPC fit but that the HPC
parametrization falls nevertheless fairly close to the data. Notice
that of the data sets shown in Fig. 4, the one for Sn/C gives the most
stringent constraint for any fit, due to the smallest error bars.

In Fig. 5, we have plotted the $Q^2$ dependence of the ratio $F_2^{\rm
Sn}/F_2^{\rm C}$ at fixed values of $x$ corresponding to those in the
NMC data \cite{NMC96-2}. The EKS98 (with CTEQ5L distributions) is shown by 
the solid lines, HIJING by the dotted lines and HPC with the dashed lines. 
The data is shown by the open squares with (statistical) error bars.
At small values of $x$ the $Q^2$ dependence is not a negligible effect.

\begin{figure}[hbt]
\vspace{-0.5cm}
\centerline{
\epsfxsize=10.5 cm\epsfbox{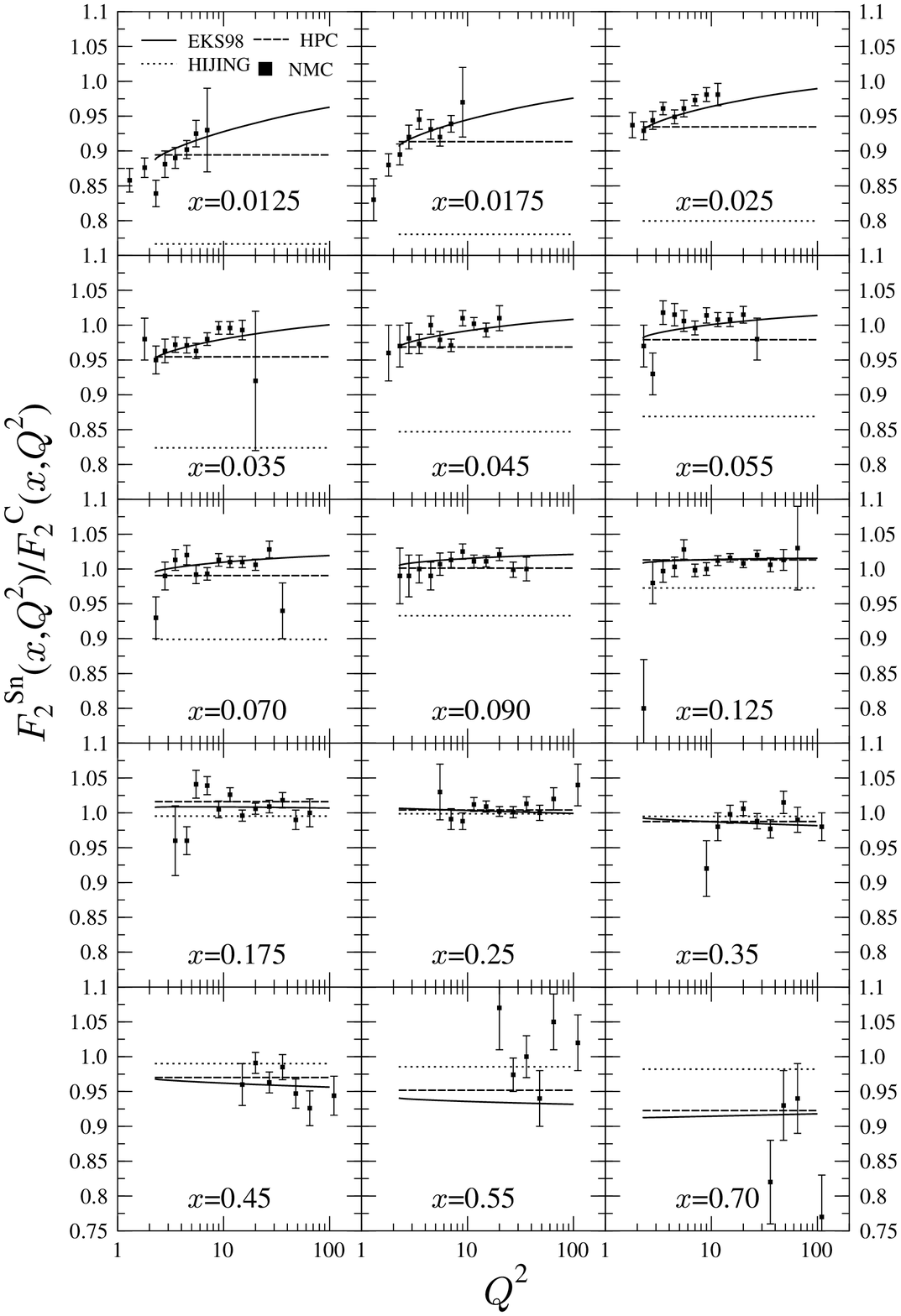}}
{ \small Figure 5. The $Q^2$ dependence of the ratio 
$F_2^{\rm Sn}/F_2^{\rm C}$ for fixed values of $x$.
The EKS98 results are the solid lines, HIJING parametrization the dotted 
line  and the HPC parametrizaton the dashed line. The NMC data \cite{NMC96-2}
is shown by the filled circles with (statistical) error bars.}
\vspace{-0cm}
\end{figure}

Finally, in Fig. 6 we show the comparison of EKS98, HIJING and HPC
with the E772 DY data \cite{E772} in $pA$ collisions. In the
computation of the ratio from Eq. (\ref{RDY}), the CTEQ5L
distributions \cite{CTEQ5L,PDFLIB} of the free proton have been used.  
The scales $Q^2$ for each $x_2$ used in computing the
EKS98 results are those in Fig. 1.  Also here the conclusion is that
the HIJING parametrization clearly overestimates the $A$ dependence of
shadowing.

\begin{figure}[tb]
\vspace{-0.5cm}
\centerline{
\epsfxsize=13cm\epsfbox{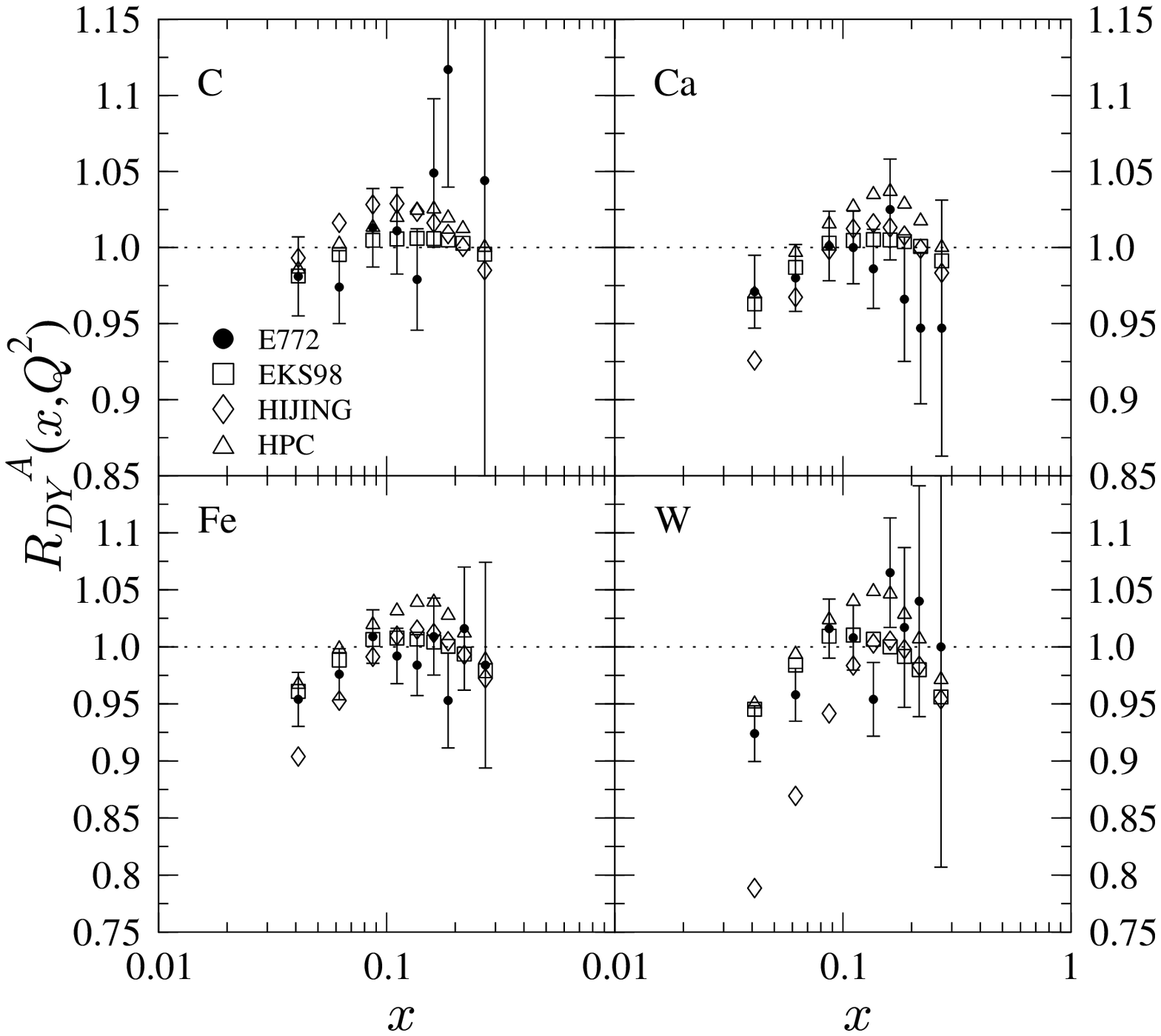}}
\vspace{0cm}
{ \small Figure 6. The Drell-Yan ratio $R_{DY}^A(x_2,\langle
Q^2\rangle)$ computed from Eq. (\ref{RDY}). The E772 data \cite{E772} 
are shown by the filled circles with error bars, the EKS98 results by 
the open squares, the HIJING parametrization by the open diamonds, 
and the HPC parametrization by open triangles.
}
\vspace{0cm}
\end{figure}

\section{Conclusions}

In this note we have discussed the determination of the nuclear parton 
distributions in the lowest order leading twist DGLAP framework. 
We have shown which kinematical range and which combinations of the 
parton distributions are probed in the measurements of the deeply 
inelastic $lA$ scatterings and the Drell-Yan process in $pA$ collisions, 
and to what extent the measurements offer constraints for the input 
distributions. The remaining uncertainties are also discussed. 

To demonstrate the differences between the different parametrizations
of the nuclear effects used in the literature, we have compared the
EKS98, HIJING and HPC parametrizations with each other and, most
importantly, with the data.  The HIJING parametrization
for $R_{F_2}^A$ clearly overestimates the $A$ dependence of nuclear
shadowing, and leads to a contradiction with the data.  Our conclusion
therefore is that for detailed studies of shadowing effects the HIJING
parametrization \cite{HIJING} is not an adequate description of the
nuclear modifications of quark and antiquark distributions. If the
flavour and scale dependence of the nuclear effects could be neglected
for some applications, we note that the HPC parametrization gives a
better representation of the data for the ratios $R_{F_2}^A$ than the
one in \cite{HIJING}. For such a use the HPC fit should, however, be
redone by including the extensive $A$ systematics provided by the NMC
measurements \cite{NMC96-1}: especially the data on the ratio
$F_2^{\rm Sn}/F_2^{\rm C}$ offers an additional constraint for the
fit.

In the leading twist framework also the conservation of baryon number
and momentum should be required, since all the nuclear effects are
contained in the parton distributions. If so done, the modifications
for the valence quarks, sea quarks and gluons differ from each other,
as is demonstrated in the DGLAP analyses
\cite{STRIKMAN,KJE,EKR,EKS98,KUMANOnew}.  If the HIJING or HPC
parametrizations of $R_{F_2}^A(x)$ are directly used to modify the
distribution of all parton flavors, the baryon number will be 
underestimated for $A=208$ by 18...24 \% with HIJING and 5...12 \% with HPC
at scales $Q^2=2.25...10^4$ GeV$^2$.

The NMC measurements \cite{NMC96-2} have revealed a clear $Q^2$
dependence in the ratio $F_2^{\rm Sn}/F_2^{\rm C}$, at $x\sim
0.01$. Naturally, scale independent parametrizations cannot reproduce
the observed behaviour.  In a DGLAP framework, the $Q^2$ dependence of
the ratio $F_2^{\rm Sn}/F_2^{\rm C}$ can be used to constrain the
nuclear gluon distributions, as first suggested in Ref. \cite{PIRNER}.
In the DGLAP analysis \cite{EKR} it was shown that the constraints
obtained for the gluons \cite{PIRNER} are also consistent with the
momentum sum rule, i.e. that a fairly strong antishadowing of gluons
appears at the input scale $Q_0^2\sim 2$ GeV$^2$. The HIJING
parametrization underestimates the momentum sum for $A=208$ by
13...23~\% and the HPC by 8...12~\% at scales $Q^2=2.25...10^4$
GeV$^2$.  In addition, within the DGLAP analysis, the data
\cite{NMC96-2} seems to rule out the case such as the Ansatz 2 in Ref.
\cite{KJE}, where gluons would have clearly stronger shadowing than
that observed in $R^A_{F_2}$.

Finally, the analysis of Ref. \cite{EKR} which lead to the EKS98
parametrization \cite{EKS98} should be improved within the DGLAP
framework in obvious ways: the fitting procedure should be automated
for inclusion of new data sets in the future, the parameter space
should be more thoroughly explored to estimate better the
uncertainties (see also \cite{KUMANOnew}) , and next-to-leading order
analysis should be performed. We do not, however, expect the results
change very much from the EKS98 \cite{EKR,EKS98}. One should also
consider expansions of the DGLAP framework, such as an inclusion of
the recombination terms \cite{GLR,MQ} in the evolution equations
\cite{QIU, KJE}, parton saturation phenomena at small values of $x$
\cite{GLR}-\cite{AS}, and possible higher-twist effects in the cross
sections \cite{QIU_ht,FRIES}, especially at lower scales. Also input
from the different models for the origin of the nuclear effects can be
considered.  However, the comparison with the data should remain as the
key feature of the analysis, since it is (at least presently) not
possible to compute the absolute nuclear parton distributions from
first principles.

\subsection*{Acknowledgements}
We thank the members of the Hard Probe Collaboration for several
discussions on the subjects considered in this paper. The financial
support from the Academy of Finland, grant no. 773101, is gratefully
acknowledged. C.A.S. has been supported by a Marie Curie
Fellowship of the European Community programme TMR (Training and
Mobility of Researchers), under the contract number HPMF-CT-2000-01025.

\end{document}